\documentclass[12pt]{article}

\usepackage{amsmath,amstext}
\usepackage{bold-extra}
\usepackage{times}
\usepackage{geometry}
\usepackage{authblk}
\usepackage[utf8]{inputenc}
\usepackage{enumitem,amssymb}
\usepackage{ragged2e}
\usepackage{graphicx}
\usepackage{cellspace}
\usepackage{natbib}
\usepackage{longtable}
\usepackage{comment}
\usepackage{url}
\usepackage{xspace}
\newlist{thematic}{itemize}{8}
\setlist[thematic]{label=$\square$}
\usepackage{pifont}

\newcommand{\BLI}{\textit{Breakthrough Listen Initiative}\xspace}

\newcommand{\MK}{\textit{MeerKAT}\xspace} 
 
\newcommand{\tseti}{{\itshape turbo}SETI\xspace}
 
\newcommand{\Parkes}{\textit{Parkes Observatory}\xspace}

\newcommand{\GBT}{\textit{Green Bank Telescope}\xspace} 
\newcommand{\APF}{\textit{Automated Planet Finder}\xspace} 

\newcommand{\baas}{BAAS}%
\newcommand{\apj}{ApJ}
\newcommand{\apjl}{ApJL}
\newcommand{\pasp}{Publications of the Astronomical Society of the Pacific}
\newcommand{\icarus}{ICARUS}
\newcommand{\aj}{AJ}
\newcommand{\spc}{\vspace{2em}}

\begin{document}
\thispagestyle{empty}
\raggedright
\large
Astro2020 Activity, Project or State of the Profession (APC) White Paper \\
\spc
\huge
The Breakthrough Listen Search for Extraterrestrial Intelligence
\spc
\linebreak
\normalsize
\textbf{Thematic areas:} technosignatures, radio astronomy, 
planetary systems, search for life beyond the Solar System, 
activities, project \\
\spc
\textbf{Primary Author:} \\ 
Name: Dr. Vishal Gajjar
\linebreak
Institution: University of California, Berkeley
 \linebreak
Email: vishalg@berkeley.edu \\
\spc
\textbf{Co-authors:} \\
Andrew Siemion (UC Berkeley, USA; SETI Institute USA; Radboud University, The Netherlands; University of Malta, Malta), 
Steve Croft (UC Berkeley,  USA; SETI Institute, USA), 
Bryan Brzycki (UC Berkeley, USA), 
Marta Burgay (INAF, Cagliari, Italy),
Tobia Carozzi (Chalmers University, Sweden),
Raimondo Concu (INAF, Cagliari, Italy),
Daniel Czech (UC Berkeley, USA), 
David DeBoer (UC Berkeley, USA),
Julia DeMarines (UC Berkeley, USA),
Jamie Drew (Breakthrough Initiatives, USA),
J. Emilio Enriquez (UC Berkeley, USA; Radboud University, The Netherlands),
James Fawcett (University of Manchester, UK),
Peter Gallagher (School of Cosmic Physics, Dublin Institute for Advanced Studies, Ireland),
Michael Garrett (University of Manchester, UK),
Nectaria Gizani (Hellenic Open University, Greece),
Greg Hellbourg (Curtin University,  Australia), 
Jamie Holder (University of Delaware, USA),
Howard Isaacson (UC Berkeley, USA; University of Southern Queensland, Australia),
Sanjay Kudale (Giant Meterwave Radio Telescope, National Center for Radio Astrophysics, Pune, India),
Brian Lacki (UC Berkeley, USA),
Matthew Lebofsky (UC Berkeley, USA),
Di Li (Key Laboratory of FAST, NAOC, CAS, China;  University of Chinese Academy of Sciences, China),
David H. E. MacMahon (UC Berkeley, USA),
Joe McCauley (Trinity College, Dublin, Ireland),
Andrea Melis (INAF, Cagliari, Italy),
Emilio Molinari (INAF, Cagliari, Italy),
Pearse Murphy (Trinity College Dublin, Ireland;  School of Cosmic Physics, Dublin Institute for Advanced Studies, Ireland),
Delphine Perrodin (INAF, Cagliari, Italy),
Maura Pilia (INAF, Cagliari, Italy),
Danny C. Price (UC Berkeley, USA; Swinburne University of Technology, Australia),
Claire Webb (UC Berkeley, USA; Massachusetts Institute of Technology, USA),
Dan Werthimer (UC Berkeley, USA),
David Williams (UC Santa Cruz, USA),
Pete Worden (Breakthrough Initiatives, USA),
Philippe Zarka (Observatoire de Paris, Paris),
Yunfan Gerry Zhang (UC Berkeley, USA)
\pagebreak
\clearpage
\pagenumbering{arabic} 
\textbf{Abstract:}
The discovery of the ubiquity of habitable extrasolar planets, combined with revolutionary advances in instrumentation and observational capabilities, have ushered in a renaissance in the millenia-old quest to answer our most profound question about the universe and our place within it $-$ Are we alone? The \BLI, announced in July 2015 as a 10-year 100M USD program, is the most comprehensive effort in history to quantify the distribution of advanced, technologically capable life in the universe. In this white paper, we outline the status of the on-going observing campaign with our primary observing facilities, as well as planned activities with these instruments over the next few years. We also list collaborative facilities which will conduct searches for technosignatures in either primary observing mode, or commensally. We highlight some of the novel analysis techniques we are bringing to bear on multi-petabyte data sets, including machine learning tools we are deploying to search for a broader range of technosignatures than was previously possible. 

\section{Key Science Goals and Objectives}
\label{sect:key_science_goals}
The search for life beyond Earth seeks to answer one of the most profound questions humans can ask about our place in the universe $-$ Are we alone? Recent discoveries of thousands of exoplanets, including many Earth-like planets \citep{2014PASP..126..398H,2015ApJ...807...45D}, point towards abundant targets of potential interest. Experiments to scan biospheres of these exo-worlds are already in the design phase and will likely start operating in the next decade \citep{biosignatures}. It is possible that some fraction of these planets host life sufficiently advanced to be capable of communicating using electromagnetic waves. The \BLI (hereafter, BL) is a US\$100M 10-year effort to conduct the most sensitive, comprehensive, and intensive search for such advanced life on other worlds \citep{2017AcAau.139...98W, Isaacson:2017ib}. BL currently has dedicated time on three telescopes: the \GBT\ \citep[GBT;][]{MacMahon:2017we} and the \Parkes\ \citep{Price:2018bv} in the radio, and \APF\ \citep[APF;][]{Lipman:2018wv} in the optical. Additionally, commensal observations will soon begin in the radio at the \MK\ radio telescope in South Africa. 
\spc\\
Technosignature searches take place in a large, multidimensional parameter space. As the luminosity function of the putative ETI transmitter is not known, deep, long observations from single pixel receivers must be complemented by wide-field surveys conducted across a significant portion of the electromagnetic spectrum. Other unknown signal characteristics include strength, intermittency, polarization, and modulation types \citep{Tarter:2003p266, 2018AJ....156..260W}. Narrow-band ($\sim$ Hz) radio signals are one of the most common signal types searched for in radio SETI \citep{Drake:1961bv, Verschuur1973, Tarter:1980p1516, 1986Icar...67..525H, Horowitz:1993p1523, NASA:2003p185, 2013ApJ...767...94S}. Such signals are ubiquitous features of early terrestrial communication systems, transit the interstellar medium easily, are readily distinguished from natural astrophysical sources and can be produced with relatively low energy \citep{1959Natur.184..844C}. The BL team has also developed a software package to search for these narrow-band signals, \tseti\footnote{\url{turboSETI: https://github.com/UCBerkeleySETI/turbo_seti}}\citep{enriquez2017apj}. Broad-band radio emission exhibiting artificial dispersion and/or underlying modulation represents a different class of radio emission indicative of an artificial origin. Due to the complexity of such signals, very limited searches have been conducted \citep[e.g.][]{2015arXiv150600055H}. Recent advances in the performance of Graphical Processing Units (GPUs) and availability of resources to store large datasets are now allowing a more effective search for these wide variety of signals (\S~\ref{sect:ML}). 
\spc\\
The primary targets of the BL program include 1 million nearby stars, 100 nearby galaxies, and deep observations of the Galactic center and plane \citep{Isaacson:2017ib}. \cite{enriquez2017apj} reported observations at 1.4\,GHz towards 692 stars using the GBT for narrowband signals. More recently, \cite{price2019apj} observed 1327 stars from $1.1  - 3.4$\,GHz using the GBT and Parkes for similar narrowband signals. Over the next five years, MeerKAT will continue to perform a survey of 1 million nearby stars (\S~\ref{sect:meerkat}). Parkes has already completed a first scan of the visible part of the Galactic plane (Price et al., 2019b in prep.). It will continue to scan the Galactic plane to investigate any intermittent ETI signals. The BL program is also planning to conduct deep observations of the Galactic center across $0.7 - 100$\,GHz, utilizing both primary observing facilities \citep{Gajjar2019gc}, and exploring partnerships with facilities operating at higher frequencies such as the Sardinia Radio Telescope in Italy (\S~\ref{sect:future_SRT}). 
\spc\\
BL also plans to examine Earth's technosignature and how it changes over time (using GBT, with plans to expand to Parkes and MeerKAT), by observing the Earth's radio signature reflected from the Moon \citep{2019BAAS...51c.558D}, following a similar study by \cite{1978Sci...199..377S}. We will also use APF to search for potential technosignatures in visible wavelengths, via Earthshine measurements. 

\section{Current status of primary BL observing facilities}

\subsection{APF}
The Automated Planet Finder is a 2.4-m telescope at Lick Observatory, Mt. Hamilton, California. The high-resolution Levy Spectrometer is the lone instrument on the telescope and is capable of producing optical spectra with a resolution of $\sim 100,000$, ideally suiting APF to search for artificial optical lasers. Since 2015 BL has used APF to search a sample of nearby stars, plus a representative sample of spectral types for stars within 50\,pc. The narrow wavelength nature and spatially-focused emission of optical lasers, allow detection against the broad band and isotropic emission of stellar light. 

In the next phase of BL observations on the APF, a select variety of planet candidates identified by the Transiting Exoplanet Survey Satellite (TESS) will be observed. We hope to develop the ability to observe stars without the benefit of standard on-slit guiding, using offset guiding techniques. This new mode would allow for tessellated observations of galaxies, which are larger than a single slit area covered by one pointing, and are too faint to use standard on-slit guiding. Stars currently too faint to observe (V $\sim$ 14) could also be observed using off-slit guiding, especially nearby M-dwarfs which are of special interest to the exoplanet community. 

\subsection{GBT}
\label{sect:GBT}
The Robert C. Byrd Green Bank Telescope (GBT) is a 100-m single dish radio telescope located in West Virginia, US. BL has contracted 20\% of the observing time since the beginning of 2016. It is the main telescope of the program in the Northern Hemisphere. Most observations to date have been concentrated on a set of nearby stars north of $-20^\circ$; this list of targets is described in \citep{Isaacson:2017ib}. The four main receivers used thus far are L-band (1.1-1.9\,GHz, 10\,Jy SEFD), S-band (1.68-2.65\,GHz, 10\,Jy SEFD), C-band (4-7.8\,GHz, 10\,Jy SEFD) and X-band (7.8-12\,GHz, 15\,Jy SEFD). Commissioning of Ku-band (11.0-15.8\,GHz, 15\,Jy SEFD), KFPA (18-27.5\,GHz, 20\,Jy SEFD) and other receivers is underway. 

In order to maximize the data output for each receiver, we installed a data recorder composed of hundreds of drives organized in 64 GPU-equipped compute nodes, and 7.8\,PB of on-site storage \citep{MacMahon:2017we}. This allows for up to 12\,GHz instantaneous bandwidth to be recorded.

The overall level of completion for the nearby star part of the program is about 90\%. We plan to finish this part of the program in the next few months. Another major part of the program for GBT is nearby galaxies, also described in \citep{Isaacson:2017ib}. Some galaxies have already been observed, although this part of the program will commence in full once the nearby star program has been finalized. 

\subsection{Parkes}
\label{sect:parkes}
Approximately a quarter of the annual observing time (1500\,h) on the CSIRO Parkes 64-m radio telescope has been allocated to BL, to survey nearby stars (within 50\,pc) and the Galactic plane (covering $\pm$6.5 degrees in Galactic latitude). The two main receivers in use are the 13-beam 20-cm receiver (MB, 1.2--1.6\,GHz, 30\,Jy SEFD), and the Ultra Wideband Low receiver (UWL, 0.7--4.0\,GHz, 25\,Jy SEFD). 
As at GBT, a wideband data recorder was installed by BL at the telescope \citep{Price:2018bv}. This recorder is capable of recording the voltage-level data products across the full bandwidth afforded by the receivers (up to 5.0\,Ghz).  The recorder consists of 27 GPU-equipped compute nodes, and a total of 3.1\,PB of on-site storage.  

\subsection{MWA}

A pilot program is under development at the Murchison Widefield Array in Western Australia. The
current system consists of a single compute node and a head node, configured in a similar manner to those at GBT and Parkes, but operating in a commensal mode. Voltage packets are sent over fiber from the site to the Listen machines in Perth. First light on a pulsar was achieved in early 2019 using playback of previously recorded data. Live commensal observing is anticipated to be operational shortly, initially using incoherent beamforming but moving shortly afterward to coherent beamforming and the generation of data products for input to the Listen SETI search pipeline. Development of this system involves an engaged group of scientists and engineers at Curtin University, who are also interested in the capabilities of the new hardware for fast transient follow-up and other work. Once the system is fully commissioned we envisage adding additional compute nodes to increase the available bandwidth and number of simultaneous beams.

\subsection{MeerKAT}
\label{sect:meerkat}
\label{MeerKAT2}
MeerKAT consists of 64 offset-gregorian antennas, each an effective 13.5m in diameter, with a maximum baseline of 8\,km. Each antenna has a carousel which can accommodate up to 4 different receivers. L-band receivers are currently installed on all antennas; UHF-band receivers are also being installed (currently available on most antennas). S-band receivers are undergoing testing prior to installation (two antennas are equipped with them) and there are plans to install X-band receivers in future. Digital signal processing is handled by up to 288 SKARAB boards. BL will conduct a commensal SETI program with MeerKAT to observe a million nearby stars and other selected targets in partnership with the South African Radio Astronomy Observatory (SARAO). Partnerships are also being formed with South African universities to engage students and foster faculty interest in SETI science.  

Space has been allocated for up to 144 BL computers in the Karoo Data Rack Area (KDRA) at the MeerKAT site. The first 1/8th of the BL equipment is currently undergoing commissioning (1 of 8 storage nodes, and 16 of 128 GPU nodes). The rest of the system will be installed in 2019Q4. The GPU nodes will ingest channelized antenna voltages into large buffers, beamform on targets in the field of view and run detection pipelines to search for signals of interest. Raw voltage data of selected signals of interest will be archived. The system also has access to a comprehensive suite of sensors providing information on all aspects of the array's operation, enabling the automation of many observing processes. Desired targets have already been preselected and are available in a database. While the primary objective is to survey one million nearby stars and other objects, there exists a possibility to conduct searches using incoherent beamforming, where signals would be localized after detection using voltage data in the buffer. In addition, proposals for limited amounts of primary observing time will be submitted for consideration for MeerKAT's dedicated Open Time allocations. 

\subsection{VERITAS}
\label{sect:veritas}
VERITAS, the Very Energetic Radiation Imaging Telescope Array System, is a set of four 12-m diameter atmospheric Cherenkov telescopes built for studying very-high-energy $\gamma$-rays.  The telescopes' large optical collection areas and fast cameras are suited as well to the search for the technosignature of fast optical flashes.
VERITAS began operations with the complete array in 2007 and operations are currently supported through 2022.  More than 10,000 hours of archival data can be searched for flashes within the 3.5$^\circ$-diameter field of view, as was done for Boyajian's Star \citep{2016ApJ...818L..33A}.
Dedicated 15-minute observations of targets from the {\it BL} list have begun in 2019, with 30\,h per year dedicated to the program.  

\section{Status of collaborative facilities and schedule} 
\label{sect:schedule_all}
In this section we highlight some of the collaborative facilities where BL will either deploy or has already deployed equipment to conduct
dedicated SETI experiments. Observation time on these facilities will either be obtained through submitting dedicated proposals to respective
time-allocation committees or through observing in commensal mode. 

\subsection{e-MERLIN and University of Manchester, UK}
Radio SETI has mostly been conducted from array instruments in standard beamformed mode due to the large volumes of interferometry data. However, recent development in computing and storage capacity allowed us to explore narrow-band imaging using array telescopes \citep{2017AJ....153..110G, 2018arXiv181007235G}. We plan to initially use e-MERLIN and eventually explore opportunities with Very-long-baseline interferometry (VLBI) to conduct such observations commensally with other on-going observations. A memorandum-of-understanding was signed in 2017 to collaborate with the 70-m Lovell Telescope and 6 of the 25-m e-MERLIN antennas, providing excellent frequency coverage from 150\,MHz to 24\,GHz. BL is collaborating with a team at the Jodrell Bank Observatory (JBO) which hosts the central correlator for the e-MERLIN array. We have installed one head-node, one compute-node, and one storage-node, connected to the e-MERLIN correlator. In early June 2019, first-light observations with this instrument were conducted with a 64\,MHz spectral window centered at 5681\,MHz towards Kepler-111 (Fawcett et al. 2019 in prep.). In 2018, in collaboration with the Square Kilometer Array and University of Manchester, BL organized a three-day workshop on the opportunities for wide-field radio SETI\footnote{http://www.jodrellbank.manchester.ac.uk/news-and-events/wide-field-seti-workshop/} and discussed the merits and challenges of this endeavor \citep{croft2019}.  \cite{2018arXiv181007235G} have already demonstrated the importance of wide-field radio SETI experiments using archival VLBI data with limited overall bandwidth. Our plan for the next few years is to deploy more equipment at JBO 
to build the necessary infrastructure to commensally observe for SETI along with e-MERLIN and VLBI observations covering the entire observing band. 

\subsection{NenuFAR and Nan\c{c}ay}
\label{sect:nenuFAR}

A Key Program of the NenuFAR commissioning process dedicated to SETI has been accepted by the Nan\c{c}ay Observatory Scientific Committee in May 2019. This program exploits the multicast capabilities of NenuFAR, a low frequency array in France, to commensally capture coherently beamformed voltage data from primary Key Programs targeting exoplanet systems and the Northern Celestial Pole. The voltage data will be locally stored and converted to the GUPPI RAW files used by BL to compress the data sets and run SETI algorithms on them. When appropriate, the SETI system will also be able to collect incoherently beamformed data commensally with other Key Programs, trading off the sensitivity of the instrument for its large field-of-view. Observations will start in July 2019 and run until January 2020. By-products of the SETI survey will allow for low-frequency Fast Radio Burst surveys using a detection pipeline developed at Curtin University, and high-frequency-resolution Radio Frequency Interference surveys at the Nan\c{c}ay Observatory. A primary observer program has also been mentioned in the Key Program proposal, consisting of a high sensitivity survey of the Northern hemisphere. This survey uses 3 beams steered at the meridian, covering 1.5$^{\circ}$ every 24\,h, resulting in a full survey in 60 days. 

The 100m-class Nan\c{c}ay Radio Telescope (NRT) standard correlator capabilities have recently been extended with the development of the dual-polarization ROACH-2-based ``Wideband Spectrometer for the NRT'' (WIBAR), offering higher frequency resolution ($\sim$ 16\,Hz) and bandwidth (550\,MHz). We successfully proposed a SETI pilot program to run on the NRT over the second semester of 2018, aimed at surveying 35 nearby stars ($<$ 10\,pc) extracted from the GAIA DR2 catalogue and the 27 galaxies from \citet{Isaacson:2017ib}. Each astronomical target is surveyed for 20\,min, and a narrowband transmission search is conducted over the resulting dynamic spectra (33 million frequency bins and 2 seconds integration per spectrum) to reach the sensitivity to detect emissions as faint as 200\,TW EIRP. 

\subsection{International LOFAR stations}
\label{sect:LOFAR}
The LOw Frequency ARray (LOFAR), is an aperture array radio telescope with antennas spread across Europe, operating at 10$-$90\,MHz and 110$-$250\,MHz \citep{2013arXiv1305.3550V}. BL is planning to use two of the international LOFAR stations: I-LOFAR located at Birr (Ireland) and LOFAR-SE located at Onsala (Sweden). These observations provide complementary sky coverage to MWA in the southern hemisphere. Moreover, due the large distance between the LOFAR stations, they provide a unique opportunity to do multi-site SETI which is crucial to discriminate sky-localized ETI signals from RFI. We are planning to do both targeted and wide-field commensal observations. We have already deployed head-nodes and planning to deploy more compute-nodes with GPUs. We will use tools developed for operating an international station in a stand-alone mode to record beamformed voltage stream data to  disk\footnote{\url{https://github.com/2baOrNot2ba/iLiSA}}. BL is working closely with the local teams at both facilities to develop tools to convert the voltage stream data to GUPPI raw voltage format \citep{lebofsky2019} for easily adapting most of the search algorithms developed for SETI within the BL team. 


\subsection{Sardinia Radio Telescope}
\label{sect:future_SRT}
The Sardinia Radio Telescope (SRT) is a 64-m antenna with a wide range of frequency coverage from 300\,MHz to 26\,GHz. New receivers are also being planned (and have been funded), up to 115\,GHz. As mentioned in \S~\ref{sect:key_science_goals}, one of the key science goals of the BL program is to conduct deep observations of the Galactic center. We aim to conduct observations at higher frequencies beyond 4\,GHz from the GBT. However, as the observing efficiency highly depends on the weather conditions at K-band (18$-$26\,GHz), having another facility observing the Galactic center along with the GBT is essential. SRT is ideally placed to carry out these observations. BL team members are in conversation with the local team to conduct feasibility studies to deploy the necessary computing infrastructure to carryout these observations. Our current plan is to deploy two compute-nodes before the end of 2019 and start observations in early 2020. We are also collaborating with a local pulsar team to build a search pipeline for pulsars near the Galactic center utilizing the same data. 

\subsection{FAST}
FAST is the largest single-aperture radio telescope in the world and provides unprecedented sensitivity with its enormous collecting area. Moreover, FAST is also equipped with a cryogenically-cooled 19-beam receiver which provides fast survey speed and efficiency in discriminating RFI for SETI. Over the next few years, FAST will commence surveys for HI, pulsars, spectral lines, VLBI, and SETI \citep{Li:2018}. In 2016 BL signed a Memorandum of understanding with the National Astronomical Observatory of China for collaboration with FAST\footnote{\url{https://breakthroughinitiatives.org/news/6}}. \cite{Li2019SETI} discuss two unique SETI experiments FAST can conduct to push limits placed by earlier studies and complement other on-going SETI activities in collaboration with BL. These surveys include targeted observations of nearby stars with exoplanets discovered from TESS. \cite{Li2019SETI} also discussed a survey of nearby galaxies, with an example covering one trillion stars in the Andromeda galaxy in just 6\,h of observing with FAST's 19-beam receiver. BL is in conversation with a local team to conduct feasibility studies early next year for a dedicated BL SETI backend.

\subsection{GMRT}
The upgraded Giant Meterwave Radio Telescope (uGMRT) is one of the most sensitive radio telescopes currently operational at mid-radio frequencies ($300 - 900$\,MHz). The long baseline elements of the GMRT provide a unique opportunity to discriminate sources of potential interference from any putative ETI signals. With the recent upgrade, GMRT has opened a new radio window along with also providing excellent sensitivity \citep{Gupta:2017fu}. It is likely that only future telescopes such as SKA1 will be able to outperform the uGMRT at centimeter wavelengths. We have proposed a pilot program to observe ten nearby stars, which host Earthlike exoplanets. BL is in conversation with the local team for a feasibility study to deploy a dedicated SETI backend. Our goal is to conduct a more ambitious program by mid-2020 involving both commensal and primary-user observing. 

\subsection{The Cherenkov Telescope Array}
The Cherenkov Telescope Array (CTA), a successor to VERITAS (\S~\ref{sect:veritas}), is a next generation very-high-energy $\gamma$-ray observatory, planned to have an array of 99 telescopes in the south (ESO, Paranal, Chile) and 19 in the north (ORM, La Palma, Spain).  Three sizes of telescopes, from 4\,m to 23\,m diameter, will be used to optimize $\gamma$-ray energy coverage of 20\,GeV to 300\,TeV.  
With the implementation of suitable triggers, CTA has extraordinary potential for technosignature searches. A single, point-like flash illuminating a whole array with uniform intensity would be an exceptional event. The 23-m telescopes, with $\sim$4 times the light collection capability of VERITAS, would be sensitive to fainter flashes. The large multiplicity of telescopes and long baselines between them would improve noise rejection and parallax measurements, respectively, compared to VERITAS.
CTA construction is planned to begin at both sites in 2020. 
The completion of the first phase of construction, with partial arrays, is planned for the beginning of 2025. The CTA instrument, plans and timeline are described by \citet{CTA_astro2020}.

\section{Technical Overview}

\subsection{Storage and Public Data}

The large size of BL datasets means that network infrastructure poses a significant challenge to egress of data from the telescopes, long term archival, and public availability of data. Nevertheless, BL has made available $\sim 1$\,PB of data in a public archive\footnote{\url{https://breakthroughinitiatives.org/news/25}}, along with tools for data analysis\footnote{\url{https://github.com/UCBerkeleySETI/turbo_seti}}\textsuperscript{,}\footnote{\url{https://github.com/UCBerkeleySETI/blimpy}}. By engaging the academic and technical communities, we can bring a range of skills to bear on our data analysis challenges, as well as helping to train the next generation of SETI scientists. In-house solutions for public data availability are currently less expensive than commercial cloud offerings, but we are piloting availability of small amounts of public data through the Google Cloud Platform, which provides easier access to machine learning tools such as TensorFlow.

\subsection{Machine Learning}
\label{sect:ML}

\subsubsection{Supervised learning for signal detection}
The first goal in radio frequency SETI is the detection of signals. When the
characteristics of a target signal are known, a labeled training set can be
created to instruct the model explicitly. In the case of rare signals, such as
fast radio bursts, where fewer detections have been made than are necessary
to construct a labeled training set from real data, simulation can be used
instead. We have already been successful at the application of classical, wide feature and deep neural network ML methods to astronomical data in our detection and classification of fast radio bursts \citep{zhang2018FRB}. These methods are directly applicable
to searching and classifying other SETI signal models, and will be
incorporated into our standard pipelines.

\subsubsection{Self-supervised Learning for Anomaly Detection}
In addition to detecting known signals, we are also interested in
signals that appear anomalous relative to the expected
instrumental and astrophysical background. For narrowband signals,
the large variability in the signals and the abundance of interfering
signals in the background data make the simulation approach unsuitable. Instead, we deployed a predictive model of anomaly detection that,
when given an observation of a narrowband signal, predicts
the next frame of observations. If the prediction deviates significantly from
the observation, an anomaly alert is triggered. Our model, based on a novel
generative adversarial architecture, is proving to be more robust than traditional
methods in detecting discrepancies between an `ON' and `OFF'
frame \citep{zhang:anomaly}.

\subsubsection{Semi-supervised Learning for Signal Classification}
The most comprehensive searches for technosignatures should probe for all
signal types that deviate from the expected natural astrophysical and instrumental
backgrounds. To address this, we would like to build a universal
signal classifier that identifies all signals, rather than just a few types that
are postulated to be more likely due to ETI. The sheer number of possibilities
in communication signals makes it difficult to construct a completely
labeled dataset for classification. In such cases a classifier for a few signals
can be built, and the other signals can be observed to fall into clusters in
the abstract representation of the model. We are experimenting and refining
this approach using software defined radios (SDR) and recorded baseband
data.



\section{Recommendations}
\subsection{Technosignatures are also biosignatures}
Technosignature searches are an integral part of astrobiology, but in the past have not been funded at the level they deserve. In recent years, Congress, and NASA, have shown renewed interest in technosignature searches --- language mandating funding for technosignature research was included in draft versions of bills in the 2018 Congress. We are strongly supportive of public funding for SETI (including public-private partnerships) which will enable SETI to become a truly integral part of the academic enterprise and an interdisciplinary area of research, opening opportunities for training and sustaining the next generation of SETI scientists and engineers. It is apparent that there is a renewed interest in the community for technosignature searches, as evident by more than half a dozen articles which were submitted for consideration for the science decadal review \citep{misraWP2019, wk2019WP, wzl+2019WP, 2019BAAS...51c.558D, 2019BAAS...51c.164L, wright2019WP, BereaWP2019, mcl+2019WP}. To bolster the notion that biosignatures are inclusive of technosignatures, the early career community has included technosignatures in their list of potential biosignatures in their publication of the Astrobiology Primer V2.0 chapter 7 (\cite{2016AsBio..16..561D}). This is noteworthy as the early career community are continuously being established in the field of Astrobiology and are emerging leaders.  

\subsection{Building an eco-system for the field}
\cite{wright2019bWP}, a whitepaper being submitted in this cycle, very clearly highlights the importance of funding to build the necessary environs for young researchers working in this area to thrive. SETI has continuously generated interest in the public and provides a genuine, and intrinsic, motivational spark for students to pursue STEM subjects in higher education. However, there is a lack of opportunity for all of the many students who want to pursue SETI as an academic career path to do so absent of federal support. As indicated by the the list of observation programs outlined above, there are plentiful opportunities for young individuals to get involved given the necessary support. In order to do that successfully, we recommend that funding should be made available to universities for researchers interested in working on the datasets being generated from various existing and upcoming facilities. 

\subsection{Interdisciplinary research for technosignature}
BL’s search for technosignatures is a scientific venture whose possibility for success holds profound consequences for all of humanity. There is much to be gained in the meantime to promote interdisciplinary research that prepares for the moment of contact. How could BL decode a message?  What assumptions does BL make regarding technical processes that might, inadvertently, underscore cultural biases? Such questions would benefit from discussion among and between SETI scientists and other scholars such as philosophers, futurists, and historians. Engagement with such questions might shape how BL teaches AI technology to search for patterns in the data, and, ethical methods on post-detection protocol. \cite{Denning2019WP}, another whitepaper being submitted with this cycle, also highlights these points. 
In that vein, we recommend to consider a sustained funding mechanism to promote interdisciplinary research that would include BL and the broader SETI community. Efforts to do so have been intermittent and decentralized—yet productive and provocative—calling attention to the need for a funded structure for continuing conversations about ET. Last May, BL hosted the Making Contact Workshop that invited historians, futurists, anthropologists, NASA policy makers, Indigenous scholars, feminist theorists, storytellers, and AI/data experts to respond to these questions: “How should SETI researchers prepare to potentially make contact with ET, and how should we respond to a signal?”  The Workshop’s goal was to collect diverse and unexpected responses with the hope to make critical connections between largely siloed scholars, and, perhaps, nudge SETI science toward greater inclusiveness. Social scientists and humanists raised topics with which SETI astronomers loosely engage, but from perspectives that digressed from the day-to-day processes of SETI science. A very brief sampling of topics raised were: the “discovery” of ET as an extended process of detection, interpretation and understanding; using practices of “radical otherness” to approach thinking through ET; the pitfalls of analogies of “contact,” and, the continuing problems of speciesism, essentialism and anthropocentrism. 
These interventions are a start on building bridges between scholars of disciplines beyond observational astronomy. The question at the heart of SETI research—Are we alone?—is, ultimately, about a question about the nature of humanity. This critical work sparked by BL is fundamental not only to the project’s success, but about knowledge-making processes themselves.

\clearpage
\setcounter{page}{1}
\pagebreak
\bibliographystyle{aasjournal}


\end{document}